\documentclass[iop]{emulateapj}
\submitted{Astrophysical Journal, accepted}

\def\teff{$T_{\rm eff}$}
\def\logg{log $g$}
\def\c12{$^{12}$C/$^{13}$C}
\def\bootes{Bo\"{o}tes I}

\begin{document}

\title{The [Fe/H], [C/Fe], and [$\alpha$/Fe] distributions of the Bo\"{o}tes I Dwarf Spheroidal Galaxy\altaffilmark{6}}

\author{David K. Lai{\altaffilmark{1,2}}, 
Young Sun Lee{\altaffilmark{3}},
Michael Bolte{\altaffilmark{1}},
Sara Lucatello{\altaffilmark{4}},
Timothy C. Beers{\altaffilmark{3}},
Jennifer A. Johnson{\altaffilmark{5}}, 
Thirupathi Sivarani,{\altaffilmark{6}},
Constance M. Rockosi{\altaffilmark{1}} }
\altaffiltext{1}{University of California Observatories/Department of Astronomy and Astrophysics, University
  of California, Santa Cruz, CA 95064; david@ucolick.org,
  bolte@ucolick.org, crockosi@ucolick.org}
\altaffiltext{2}{NSF Astronomy and Astrophysics Postdoctoral Fellow}
\altaffiltext{3}{Department of Physics $\&$ Astronomy and JINA: Joint Institute for
  Nuclear Astrophysics, Michigan State University, East Lansing, MI
  48824,USA; lee@pa.msu.edu, beers@pa.msu.edu}
\altaffiltext{4}{Osservatorio Astronomico di Padova, Vicolo dell'Osservatorio 5, 35122 Padua, Italy; sara.lucatello@oapd.inaf.it.}
\altaffiltext{5}{Department of Astronomy, Ohio State University, 140
  W. 18th Ave., Columbus, OH 43210; jaj@astronomy.ohio-state.edu.}
\altaffiltext{6}{Indian Institute of Astrophysics, 2nd block
  Koramangala, Bangalore 560034, India; sivarani@iiap.res.in}
\altaffiltext{7}{The data presented herein were obtained at the
  W.M. Keck Observatory, which is operated as a scientific partnership
  among the California Institute of Technology, the University of
  California and the National Aeronautics and Space A
  dministration. The Observatory was made possible by the generous
  financial support of the W.M. Keck Foundation.}

\begin{abstract}

We present the results of a low-resolution spectral abundance study of
25 stars in the \bootes{} dwarf spheroidal (dSph) galaxy. The data
were obtained with the LRIS instrument at Keck Observatory, and
allow us to measure [Fe/H], [C/Fe], and [$\alpha$/Fe] for each
star. We find both a large spread in metallicity (2.1 dex in [Fe/H])
as well as the low average metallicity in this system,
$\langle$[Fe/H]$\rangle=-2.59$, matching previous estimates. This
sample includes a newly discovered extremely metal-poor star, with
[Fe/H]=$-3.8$, that is one of the most metal-poor stars yet found in a
dSph. We compare the metallicity distribution function of \bootes{} to
analytic chemical evolution models. While the metallicity distribution
function of \bootes{} is best fit by an Extra Gas chemical evolution
model, leaky-box models also provide reasonable fits. We also find
that the [$\alpha$/Fe] distribution and the carbon-enhanced metal-poor
fraction of our sample (12\%) are reasonable matches to Galactic halo
star samples in the same metallicity range, indicating that at these low
metallicities, systems like the \bootes{} ultra-faint dSph could have
been contributors to the Galactic halo.

\end{abstract}


\section{Introduction}

In the last few years, numerous ultra-faint dwarf 
spheroidal (dSph) galaxy companions to the Galaxy have been discovered based on
the Sloan Digital Sky Survey imaging (e.g, \citealt{willman10}, and
references therein). Recent studies have shown that these galaxies contain stars
that are generally very metal poor; the mean [Fe/H] values of the ultra-faint
dSphs is consistent with an extension of the luminosity-metallicity relationship
found in their more luminous counterparts (e.g, \citealt{norris10b,kirby11a}).
Systems such as these may thus have been important contributors to the most
metal-poor populations of the Galactic stellar halo.

This paper presents a chemical abundance analysis of stars in the \bootes{} dSph
galaxy based on low-resolution spectroscopy. Originally discovered by
\citet{bootesI}, \bootes{} has been the subject of a number of previous studies. 
For example, \citet{munoz06} and \citet{martin07} measured $M/L \sim 130
- 600$, making this one of the most dark-matter-dominated galaxies in the Local
Group, a result consistent with the N-body simulations of \citet{fellhauer08}.

The metallicity of \bootes{} has been estimated by several studies as well.
\citet{bootesI} noted the similarity of its color-magnitude diagram (CMD) to
the fiducial of M92, a globular cluster with [Fe/H]$=-2.3$. Using
RR Lyrae periods, \citet{siegel06} reported a range in [Fe/H] of $\sim -2.0$
to $-2.5$. \citet{hughes08} used Washington $CT_1T_2$ photometry to
estimate [Fe/H]$=-2.1$, while \citet{martin07}, using the calcium triplet (CaT)
method, obtained a mean metallicity of [Fe/H]$=-2.1$ based on a sample
of 30 stars. However, it has been shown that previous calibrations of the 
CaT estimator are unreliable at lower metallicities
(e.g., \citealt{battaglia08,kirby08a}). \citet{norris08} find a mean
[Fe/H] of $-2.51$ for a sample of 16 stars, using the \ion{Ca}{2} K line-strength
index approach.

High-resolution spectroscopic studies of \bootes{} members have also been
carried out. \citet{norris10} presented a study of a member star with
[Fe/H]$=-3.66$, and \citet{norris10b} updated their previous work with
high-resolution determinations of [Fe/H] and [C/Fe] for six stars (as
well as deriving [C/Fe] from their original moderate-resolution
data). \citet{feltzing09} presented a high-resolution spectroscopic study
for seven \bootes{} stars, finding one star with an anomously high
[Mg/Ca] ratio as compared with the rest of their sample. This signature
is rarely found in metal-poor halo stars \citep{aoki07b,cohen07},
However, this may be a more common feature among dSph stars, as
evidenced from abundance studies of stars in Hercules \citep{koch08}
and Draco \citep{fulbright04}.

In this study we present an abundance analysis for 25 stars in the
\bootes{} system, based on low-resolution spectra obtained with the
LRIS multi-object spectrometer on the Keck I telescope. Section 2
summarizes details of these observations and reductions. Section 3
describes an automated analysis approach used to determine estimates
of [Fe/H], [$\alpha$/Fe], and [C/Fe] for each star. In Section 4 we
combine our results with those of \citet{feltzing09} and
\citet{norris10b}, comparing the \bootes{} metallicity distribution to
both other dSph galaxies and with chemical evolution models, and also
the [$\alpha$/Fe], and [C/Fe] distributions to the stellar abundances
of Galactic halo stars.

\section{Observation and Reduction Details}

Our \bootes{} targets were selected from the radial velocity study of
\citet{martin07}. The stars with confirmed radial velocity membership
in \bootes{} were observed using the blue side of the LRIS
multi-object spectrometer on the Keck I telescope
\citep{oke95,steidel04}. For all of our targets we used the 600/4000
Grism, with a slit width of 0.7'', yielding a resolving power of R
$\sim$ 1800 at 5100 \AA{}. We were able to observe 4 masks, giving us
a total of 25 stars in the \bootes{} system. The stars range from near
the center to $\sim8$' of center, all inside of the \bootes{} half-light
radius of 13' \citep{bootesI}. The only criteria used in
selecting Boo members was the radial velocity membership from
\citet{martin07}, so our sample should provide an unbiased view of the
nature of the stars in this galaxy. We also observed one mask each
for the globular clusters M5 and M15, in order to validate our
abundance measurements and characterize their uncertainties. There is
no widely accepted naming scheme for the stars in \bootes{}, so we
employ the scheme from Table 1.

Standard IRAF\footnote{IRAF is distributed by the National Optical Astronomy Observatories,
    which are operated by the Association of Universities for Research
    in Astronomy, Inc., under cooperative agreement with the National
    Science Foundation.} routines were used to remove instrumental signatures from the data, extract 
spectra from the two-dimensional images, and wavelength calibrate the spectra.  
Due to flexure in LRIS, we could not obtain reliable independent radial
velocity estimates from our measurements.

\begin{deluxetable*}{lccccccrr}
\tablecolumns{9}
\tabletypesize{\small}
\tablecaption{Coordinates and Abundance Results for \bootes{} Program Stars \label{infotable}}
\tablewidth{0pc}
\tablehead{
\colhead{Star} & \colhead{R.A.} & \colhead{Decl.} & \colhead{SNR} & \colhead{\teff{}} & \colhead{\logg{}} & \colhead{[Fe/H]} & \colhead{[$\alpha$/Fe]} & \colhead{[C/
Fe]} \\
\colhead{ID} & \colhead{(J2000.0)} & \colhead{(J2000.0)} & \colhead{(@5180\AA{})} & \colhead{(K)} & \colhead{(cgs)} & \colhead{(dex)} & \colhead{(dex)} & \colhead{(dex)} 
}
\startdata
    Boo01       &  14 00 10.49 & +14 31 45.50 &    130  &  4716 &      1.65 &   $-$2.34 &      0.12 &   $-$0.50 \\
    Boo02       &  14 00 12.92 & +14 33 11.80 &     47  &  5114 &      2.00 &   $-$2.37 &      0.28 &      0.36 \\
    Boo03       &  14 00 33.08 & +14 29 59.70 &     45  &  5127 &      1.99 &   $-$3.09 &      0.43 &      0.79 \\
    Boo04       &  14 00 03.08 & +14 30 23.60 &     25  &  5210 &      2.68 &   $-$2.39 &      0.35 &      0.34 \\
    Boo05       &  13 59 44.27 & +14 32 41.20 &     18  &  5077 &      2.55 &   $-$2.89 &      0.05 &   $<$0.00 \\
    Boo06       &  13 59 52.33 & +14 32 45.70 &     23  &  5404 &      2.40 &   $-$2.20 &      0.32 &      0.64 \\
    Boo07       &  14 00 05.34 & +14 30 23.30 &     38  &  5200 &      2.54 &   $-$2.49 &      0.16 &      0.39 \\
    Boo08       &  13 59 44.96 & +14 32 30.10 &     22  &  5178 &      2.30 &   $-$2.48 &      0.18 &      0.32 \\
    Boo09       &  14 00 23.38 & +14 32 45.30 &     13  &  5563 &      2.44 &   $-$2.65 &      0.37 &      0.90 \\
    Boo10       &  13 59 51.08 & +14 30 49.80 &     11  &  5086 &      2.42 &   $-$2.59 &   \nodata &   $<$0.00 \\
    Boo11       &  13 59 50.76 & +14 31 14.20 &     14  &  5199 &      2.65 &   $-$2.43 &      0.05 &   $-$0.05 \\
    Boo12       &  14 00 27.29 & +14 32 19.60 &     44  &  5168 &      2.19 &   $-$2.48 &      0.04 &      0.24 \\
    Boo13       &  14 00 22.45 & +14 33 26.90 &     41  &  5631 &      2.35 &   $-$2.49 &      0.23 &   $<$0.00 \\
    Boo14       &  13 59 48.34 & +14 32 03.60 &     32  &  5971 &      2.56 &   $-$2.57 &      0.34 &   $<$0.00 \\
    Boo15       &  13 59 57.85 & +14 28 02.50 &     23  &  5117 &      2.31 &   $-$2.89 &      0.30 &   $<$0.00 \\
    Boo18       &  14 00 03.33 & +14 28 51.50 &     22  &  5287 &      2.27 &   $-$2.51 &      0.13 &      0.44 \\
    Boo19       &  14 00 05.61 & +14 26 18.90 &     35  &  5141 &      2.38 &   $-$3.29 &      0.28 &      0.40 \\
    Boo20       &  13 59 47.07 & +14 28 52.60 &     12  &  4931 &      2.44 &   $-$2.42 &      0.13 &   $-$0.35 \\
    Boo21       &  14 00 09.85 & +14 28 23.00 &    125  &  4775 &      1.48 &   $-$3.79 &      0.27 &      2.20 \\
    Boo22       &  13 59 50.63 & +14 29 11.10 &     36  &  4866 &      1.93 &   $-$2.87 &      0.12 &   $-$0.16 \\
    Boo23       &  14 00 11.54 & +14 25 56.10 &     20  &  5475 &      2.83 &   $-$2.90 &      0.16 &      1.86 \\
    Boo24       &  14 00 25.83 & +14 26 07.60 &     80  &  4798 &      1.63 &   $-$1.65 &      0.46 &   $-$0.80 \\
    Boo25       &  14 00 21.84 & +14 25 53.40 &     24  &  5141 &      2.02 &   $-$2.76 &      0.25 &      1.34 \\
    Boo28       &  14 00 02.29 & +14 26 53.40 &     20  &  5449 &      3.29 &   $-$2.31 &      0.06 &      0.27 \\
    Boo30       &  14 00 23.34 & +14 26 08.00 &     17  &  5449 &      3.29 &   $-$1.86 &      0.51 &      0.38 
\enddata
\end{deluxetable*}

\section{Stellar Parameters and Analysis \label{analysis}}

\begin{figure*}
\begin{center}
\scalebox{.7}[.7]{ \plotone{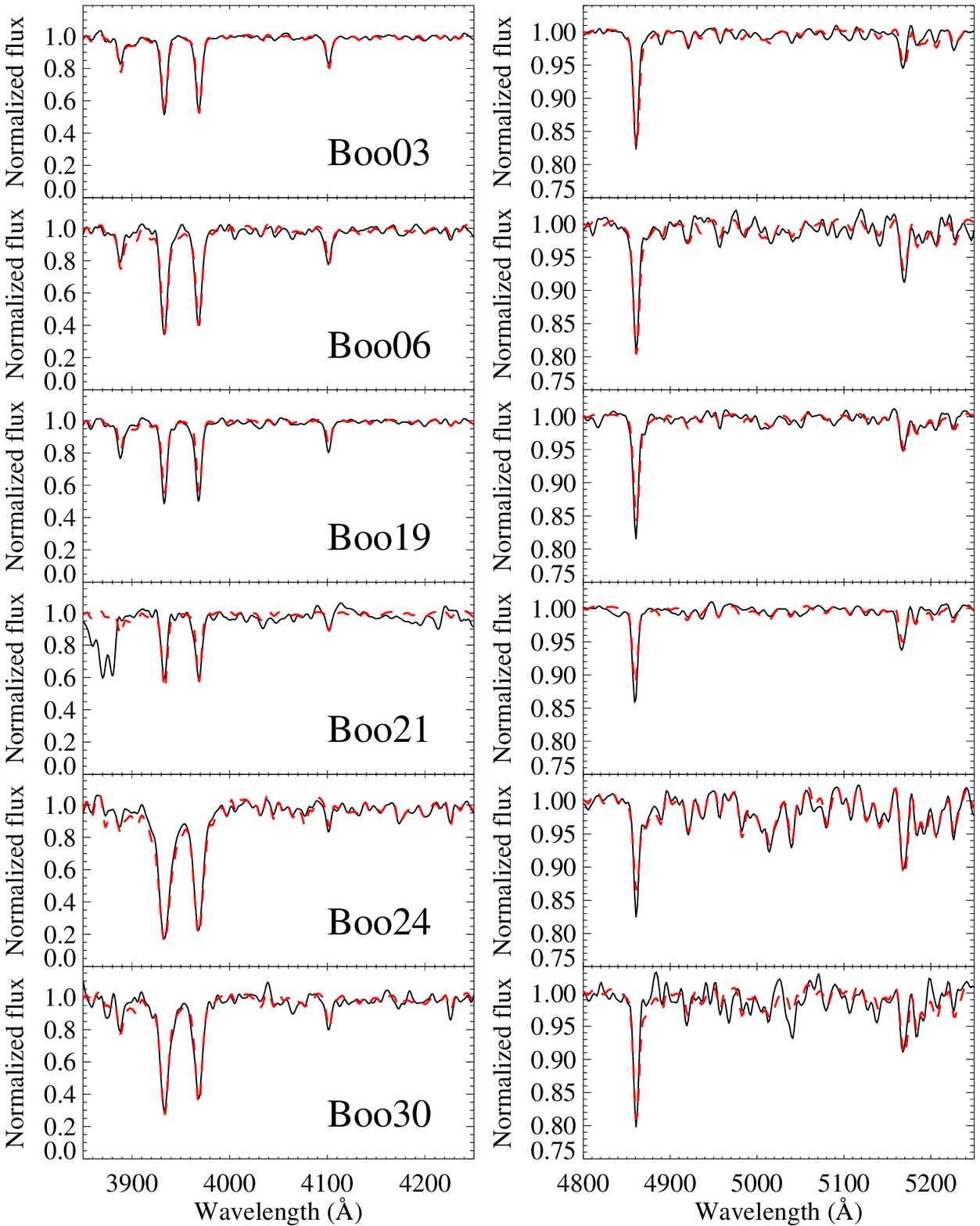}}
\end{center}
\figcaption{Examples of the observed spectrum (solid lines) along with
  synthetic spectrum (dashed lines) produced using their respective
  n-SSPP derived stellar parameters. Two regions are shown for each
  star to highlight the Ca II H \& K region and the Mg triplet
  region. The synthetic spectra are in very good agreement with the
  observed spectra. The very high [C/Fe] (and possibly [N/Fe]) of
  Boo21 is reflected in the poor fit of the CN 3880 feature, as the
  n-SSPP does not estimate carbon or nitrogen and the synthetic
  spectra are produced with the assumption of solar abundance ratios
  for carbon and nitrogen. 
\label{fit1}
}
\end{figure*}

We employed a newly developed version of the SEGUE Stellar Parameter
Pipeline (SSPP; \citealt{lee1,lee2,allende08}), called the n-SSPP
(indicating its use for non-SEGUE data). This code is suitable for
application to spectra other than those taken by the SDSS/SEGUE that
are of approximately the same resolution (R$\sim 2000$), and cover a
wavelength range of (ideally) at least $\sim$3900 to 5500 \AA{}. In
conjuction with the spectra, the n-SSPP also uses $ugriz$ colors
when available (as is the case with our program stars). Otherwise it
makes use of Johnson $V$ magnitude and $B-V$ colors, and/or a 2MASS
\citep{2mass} $J$ magnitude and $J-K$ color, when available, in order
to predict magnitudes and colors on the SDSS photometric system. The
n-SSPP then determines the primary atmospheric parameters (\teff{},
\logg{}, [Fe/H], and [$\alpha$/Fe]), based on a subset of the
procedures described in \citet{lee1} and
\citet{lee2010}. \citet{lee2010} describes the calibration and
validation of the techniques used for estimating [$\alpha$/Fe], to
which we refer the interested reader. Examples of observed spectra
matched to synthetic spectra produced with the n-SSPP derived
parameters are shown in Figure \ref{fit1}.

\begin{figure*}
\begin{center}
\scalebox{.85}[.85]{ \plotone{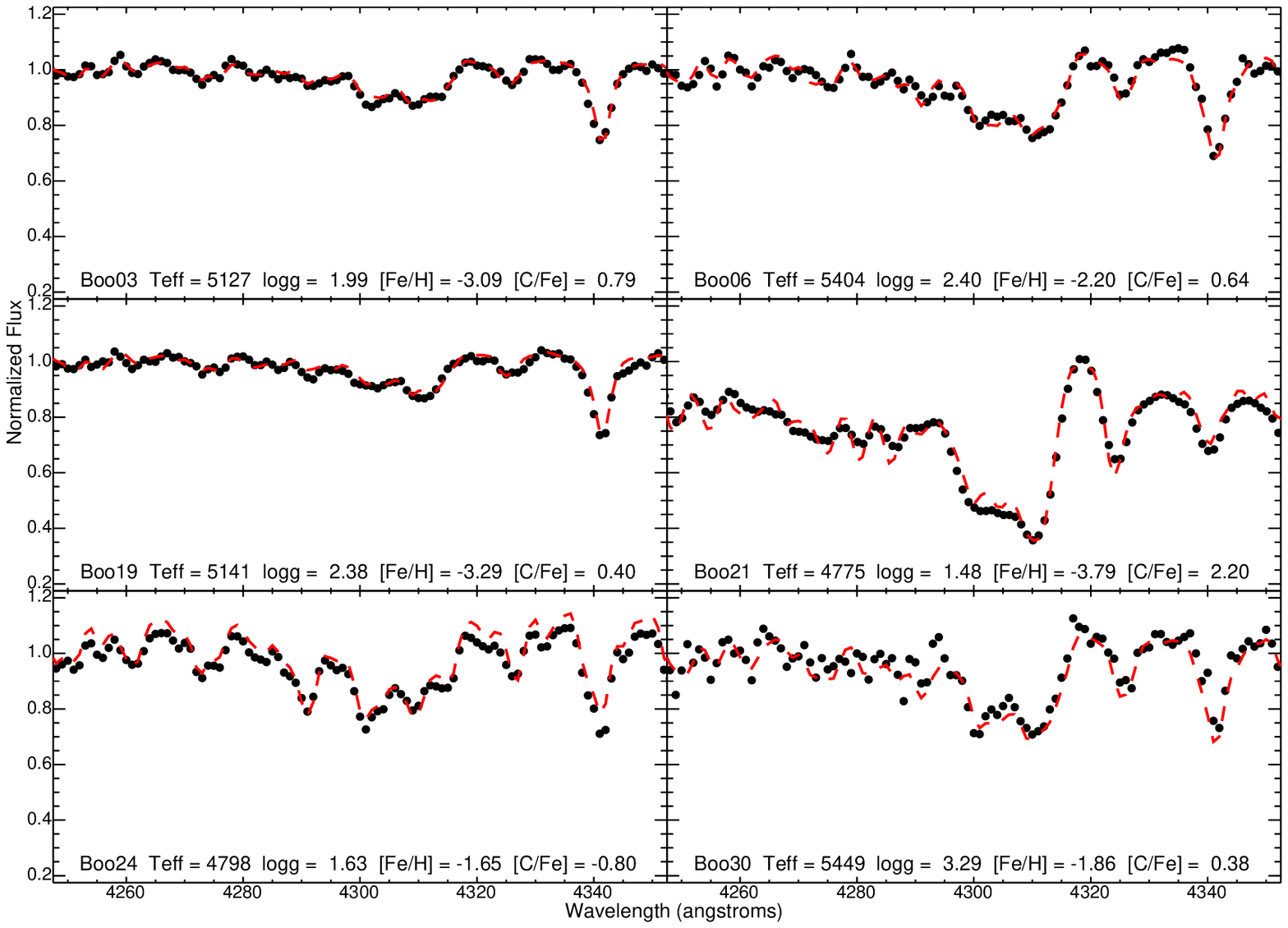}}
\end{center}
\figcaption{Examples of synthetic spectra (dashed lines) used to
  determine the [C/Fe], as compared to the observed spectra (filled
  circles) of the same stars shown in Figure \ref{fit1}. 
\label{fit2}
}
\end{figure*}

Carbon abundances, with accuracies on the order of 0.3 dex, were
estimated from the CH G band at $\sim 4300$\AA{} by matching the
observed spectra near this feature with an extensive grid of synthetic
spectra. The determination of carbon abundance can be affected by the
oxygen abundance because of molecular equilibrium. Since no
independent [O/Fe] can be determined from our spectra, the grid
assumes [O/Fe]$ = +0.4$ for metallicities below solar, and [O/Fe]$
=0.0$ for solar and above. In Figure \ref{fit2}, we show examples of
the synthetic fits of the G band region used to determine [C/Fe]
ratios. Details of this procedure are described by
\citet{beers07b}, as extended by \citet{carollo11}. The final adopted
stellar parameters and abundances are presented in Table
\ref{infotable}.

\subsection{Comparisons to Previous Results: Verification of the Pipeline and Accuracy Estimates}

\subsubsection{[Fe/H] comparisons}

The SSPP has already been verified to work well with SDSS spectra
\citep{smolinski11}. However, we also
checked the accuracy of the n-SSPP pipeline using our particular setup, with
observations of stars in two globular clusters M5 and M15. Table
\ref{clusterinfo} lists the atmospheric parameters and abundances derived using
the n-SSPP for
stars from the two clusters. Table \ref{infotable} and
\ref{clusterinfo} also list the final signal-to-noise ratios (SNRs) of
the spectra as estimated from the total counts of each extracted spectrum
(at 5180\AA{}). While it was not possible to exactly match the SNR range between the
cluster data and the \bootes{} stars, due to varying observing conditions and
magnitude differences, the SNR range is comparable between the cluster spectra
and the \bootes{} spectra.

\begin{deluxetable*}{lcccccccccccc}
\tablecolumns{11}
\tabletypesize{\small}
\tablecaption{Abundance Results for M5 and M15 \label{clusterinfo}}
\tablewidth{0pc}
\tablehead{
\colhead{Star} & \colhead{SNR} & \colhead{\teff{}} & \colhead{\logg{}} & \colhead{[Fe/H]} & \colhead{[$\alpha$/Fe]} & 
\colhead{[C/Fe]} & \colhead{\teff{}} & \colhead{\logg{}} & \colhead{[Fe/H]} & \colhead{[$\alpha$/Fe]}\\
\colhead{ID} & \colhead{(@5180\AA{})} &\colhead{(K)} & \colhead{(cgs)} & \colhead{(dex)} & \colhead{(dex)} & \colhead{(dex)} &
\colhead{lit.} & \colhead{lit.} & \colhead{lit.} & \colhead{lit.}
}
\startdata
  M5 I-2   &   30  & 4359 &      2.06 &   $-$1.36 &   \nodata &   $<$0.00   &   4540   &  1.52    & $-$1.15   &    0.31   \\
  M5 I-4   &   35  & 4638 &      2.22 &   $-$1.19 &      0.33 &   $-$0.27   &  \nodata &  \nodata &   \nodata &   \nodata \\
  M5 I-50  &   29  & 4581 &      2.39 &   $-$1.25 &      0.29 &   $-$0.66   &   4590   &  1.55    & $-$1.18   &    0.28   \\
  M5 I-58  &   37  & 4571 &      2.10 &   $-$1.01 &      0.09 &   $-$0.72   &   4400   &  1.60    & $-$1.10   &    0.28   \\
  M5 I-61  &   35  & 4567 &      1.89 &   $-$1.36 &      0.23 &   $-$0.39   &   4425   &  1.27    & $-$1.17   &    0.28   \\
  M5 I-68  &   50  & 4000 &      1.81 &   $-$1.04 &   \nodata &   $<$0.00   &   4066   &  0.63    & $-$1.17   &    0.29   \\
  M5 I-71  &   41  & 4373 &      2.05 &   $-$1.01 &   \nodata &   $-$0.52   &   4360   &  1.12    & $-$1.08   &    0.26   \\
  M5 II-50 &   29  & 4980 &      2.23 &   $-$0.88 &      0.06 &   $-$0.15   &   4590   &  1.57    & $-$1.18   &    0.24   \\
  M5 II-59 &   37  & 4541 &      2.06 &   $-$1.27 &      0.11 &   $-$0.71   &   4450   &  1.27    & $-$1.08   &    0.27   \\
  M15 K144 &   138 & 4706 &      1.15 &   $-$2.20 &      0.11 &   $-$0.52   &   4425   &  0.75    & $-$2.26   &    0.29   \\
  M15 K169 &   92  & 4527 &      1.42 &   $-$2.29 &      0.21 &   $-$0.90   &   4400   &  0.65    & $-$2.37   &    0.32   \\
  M15 K255 &   31  & 4571 &      1.43 &   $-$2.38 &      0.19 &      0.74   &   4640   &  1.40    & $-$2.30   &    0.18   \\
  M15 K341 &   37  & 4333 &      1.35 &   $-$2.04 &   \nodata &   $-$0.98   &   4275   &  0.45    & $-$2.28   &    0.51   \\
  M15 K431 &   45  & 4939 &      1.89 &   $-$2.05 &      0.29 &   $-$0.48   &   4375   &  0.50    & $-$2.36   &    0.43   \\
  M15 K553 &   53  & 4925 &      2.35 &   $-$2.43 &      0.21 &   $<$0.00   &   4855   &  2.00    & $-$2.30   &    0.27   \\
  M15 K64  &   39  & 4764 &      1.77 &   $-$2.67 &      0.30 &   $<$0.00   &   5100   &  2.25    & $-$2.30   &   \nodata \\
  M15 K875 &   20  & 5268 &      2.28 &   $-$2.09 &      0.05 &   $<$0.00   &   4775   &  1.65    & $-$2.30   &    0.40
\enddata
\end{deluxetable*}

Most of the cluster stars were selected because they had previous abundance
determinations based on high-resolution spectra. The literature information
listed in Table \ref{clusterinfo} come from the high-spectral resolution studies
of \citet{ivans01} for M5, and \citet{sneden2000}, or when available,
\citet{sneden97}, for M15 (the [Fe/H] have been adjusted using a solar log$\epsilon$(Fe)
$=7.45$, to match this study). Using these high-resolution studies as benchmarks
the M5 stars have $\langle {\rm [Fe/H]} \rangle=-1.13, \sigma=0.05$, and
the M15 stars have $\langle {\rm [Fe/H]} \rangle=-2.31, \sigma=0.04$ (the
$\langle \rangle$ notation, here and elsewhere, refers to a straight
average unless otherwise noted).
For our purposes, we assume that these values are the true metallicities for
each cluster, and that the true intrinsic spread of metallicities is no greater
than the standard deviation from the high-resolution determinations. The
respective values derived from the LRIS observations are $\langle {\rm [Fe/H]}
\rangle=-1.15, \sigma=0.17$ for M5 and $\langle {\rm [Fe/H]} \rangle=-2.27,
\sigma=0.22$ for M15. We take the much larger $\sigma$s as an indication of the
error in the [Fe/H] measurements from this study.

We have two \bootes{} stars in our sample which were previous analyzed
in the study of \citet{norris10b}. One of these was also part of the
study of \citet{feltzing09}. The [Fe/H] abundances in these two
studies are derived from analysis of high-resolution spectra. The
stars in common are Boo01 (Boo-117 in \citealt{norris10b} and
\citealt{feltzing09}) and Boo24 (Boo-041 in \citealt{norris10b}). For
Boo01, we find [Fe/H]$=-2.34$, in very good agreement with the
values of [Fe/H]$=-2.31$ and [Fe/H]$=-2.25$ reported by \citet{feltzing09} and
\citet{norris10b}, respectively. We find Boo24 to have [Fe/H]$=-1.65$,
also in reasonable agreement with the [Fe/H]$=-1.93$
value reported by \citet{norris10b} given the cluster comparison $\sigma$s.

\begin{figure*}
\begin{center}
\scalebox{.9}[.9]{ \plotone{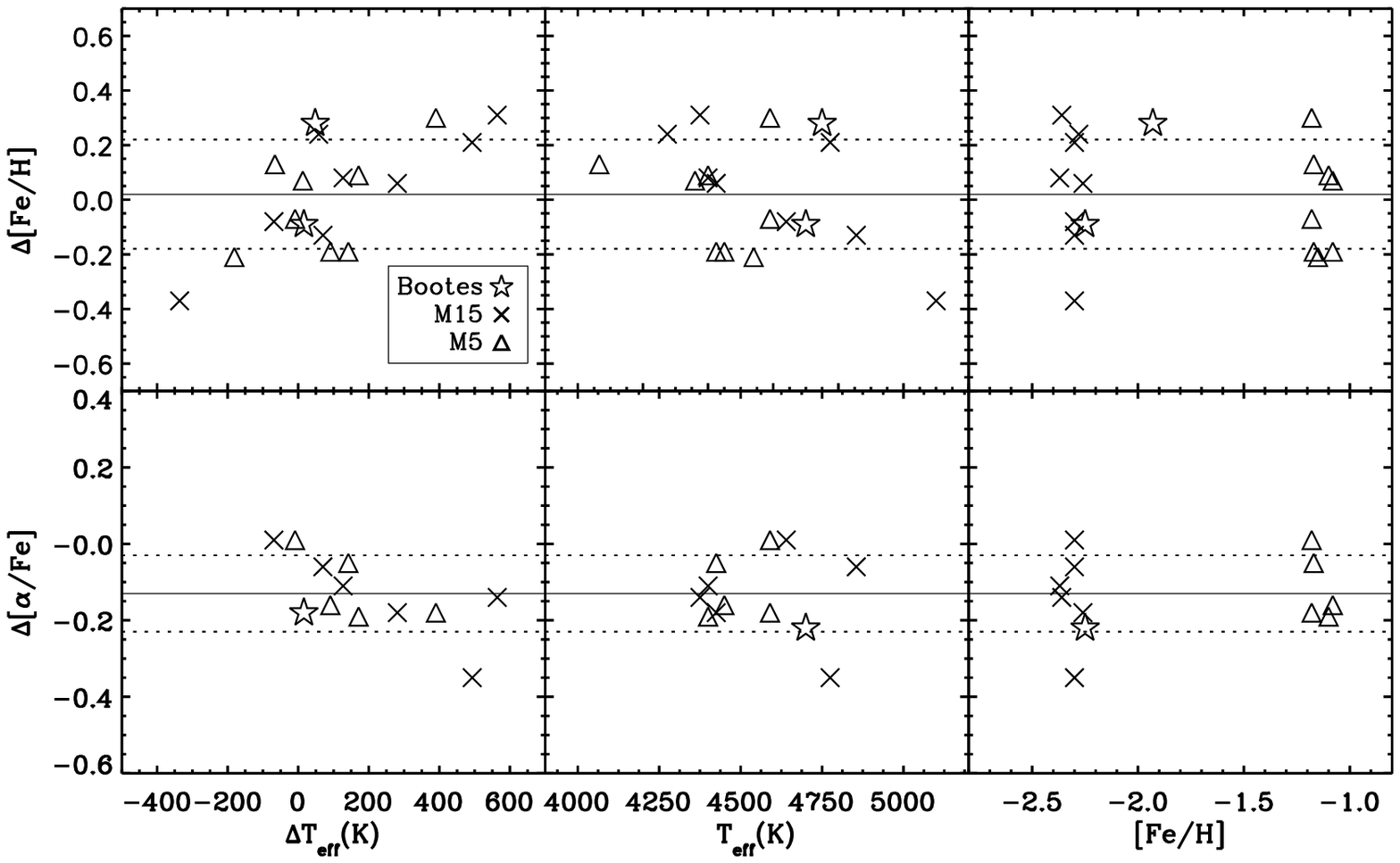}}
\end{center}
\figcaption{The top panels plot $\Delta{\rm [Fe/H]}$ as a function of
  $\Delta$\teff{} (both in the sense of our values minus the previous literature
  values), literature \teff{} values, and literature [Fe/H] values. The average
  $\langle \Delta{\rm [Fe/H]}\rangle =+0.02$ is plotted as the solid line, and
  $\sigma(\Delta{\rm [Fe/H]})=0.20$ is plotted with the dashed lines. The bottom
  panels are the same, but for $\Delta{\rm [}\alpha{\rm /Fe]}$. $\langle
  \Delta{\rm [}\alpha{\rm /Fe]} \rangle =-0.13$ is plotted with the solid line,
  and $\sigma(\Delta{\rm [} \alpha{\rm /Fe]})=0.10$ is plotted with the dashed
  lines. Out of all of these cases, only $\Delta{\rm [Fe/H]}$ as a function of
  $\Delta$\teff{} shows a significant correlation, with a probability of
  correlation $>95$\%.
\label{comp}
}
\end{figure*}

The difference between this study's [Fe/H] determinations and the
previous literature determinations ($\Delta$[Fe/H]) for the M5, M15
and \bootes{} stars is summarized in the top panels of Figure
\ref{comp}. Overall, there is only a very small ($+0.02$ dex) offset, in
the sense of this study minus literature values. Figure \ref{comp}
plots this difference as a function of 
$\Delta$\teff{} (in the sense of our values minus literature values), \teff{}, and [Fe/H]. The clear correlation
with $\Delta$[Fe/H] with $\Delta$\teff{} show that some of the
differences in [Fe/H] may be related to the adopted \teff{} values of
this study. However, there is a spread in $\Delta$[Fe/H] even when the
\teff{} values are in agreement. There is also no significant
correlation with either \teff{} or [Fe/H] of a star, indicating that
we are free from systematic offsets, at least over the range defined by
the comparison-star parameters.  Therefore, we
estimate the final random individual [Fe/H] errors as the standard
deviation of $\Delta$[Fe/H] for the entire comparison sample, $\sigma
(\Delta{\rm [Fe/H]})=0.20$. This agrees well with the individual $\sigma$s found
for the M5 and M15 stars.

\subsection{[$\alpha$/Fe] comparisons}

Figure \ref{comp} also compares the difference between determinations
of [$\alpha$/Fe] from this study with previous literature estimates 
($\Delta$[$\alpha$/Fe]). This is not an exact comparison, as the
previous literature studies measure different individual $\alpha$-elements,
including Mg, Si, Ca, and Ti. To calculate the average
literature [$\alpha$/Fe] values, we simply averaged whichever of the
$\alpha$-elements listed above were available in each star's
respective high-resolution study. While not a perfect comparison, this
gives a good handle on our errors and any possible offsets in
interpreting the \bootes{} sample.

The $\sigma$($\Delta$[$\alpha$/Fe]) from the comparisons is 0.10
dex. We take this value as the estimate on the random error of the
[$\alpha$/Fe] determinations. However, note that this study's [$\alpha$/Fe]
values are offset by an average of $-0.13$ dex from literature
measurements. A very similar offset was found when comparing
high-resolution M15 abundances from the same sources as used here to
SDSS-derived values by \citet{lee2010}, who pointed out that some of
the offset could be due to the observational errors on [Si/Fe] and
[Ti/Fe] reported in \citet{sneden97}. However, because we calculate
the average offset including the M5 and \bootes{} high-resolution
determinations, we caution that this average offset in
$\Delta$[$\alpha$/Fe] is a potential systematic error in our analysis.

\subsection{[C/Fe] comparisons}

There are no independent measurements of [C/Fe] from
high-spectral-resolution studies for any one of our individual cluster
stars. However, we can compare the $\langle {\rm [C/Fe]} \rangle$ we
derive from this study to the high spectral-resolution study of
\citet{lai11a} for 17 stars in M5. In this study, we find $\langle
{\rm [C/Fe]} \rangle=-0.49$ and $\sigma({\rm [C/fe]})=0.22$, as
compared to $\langle {\rm [C/Fe]} \rangle=-0.27$ and
$\sigma({\rm [C/Fe]})=0.25$ found by Lai et al. A caution is that [C/Fe] can
vary greatly in individual globular cluster stars, so not being able to compare
star-to-star values of [C/Fe] has its limitations. However, we can also directly
compare the [C/Fe] derived in the overlapping \bootes{} stars with the values
found by \citet{norris10b}. For Boo01 and Boo24, \citet{norris10b} derive [C/Fe]
values of $-0.30$ and $-0.65$, respectively. These compare reasonably well with
this study's values of $-0.05$ and $-0.80$ for the same stars. 

\citet{carollo11} have compared determinations of [C/Fe] employing the same techniques for
carbon abudundance estimates as the present study, but for a sample of very
low-metallicity halo stars from SDSS with high-resolution spectroscopic
determinations reported by W. Aoki et al., in preparation (covering the same
metallicity range as the \bootes{} sample). They find a very small offset
(+0.03 dex) and rms scatter of 0.3 dex from this exercise. 

In summary, we find good agreement between our [Fe/H] values and those derived
by previous high-resolution studies. A reasonable estimate of our individual
errors comes from $\sigma (\Delta{\rm [Fe/H]})=0.20$ and
$\sigma$($\Delta$[$\alpha$/Fe])$=0.10$ dex. We caution that a systematic offset
on the order $-0.13$ dex for [$\alpha$/Fe] may also be present. It is more
difficult to quantify the individual errors in [C/Fe], however the general
agreement of our values with previous studies provides confidence in the values
derived here.

\section{Results \label{discussion}}

\begin{figure}
\begin{center}
\scalebox{1.1}[1.1]{
\plotone{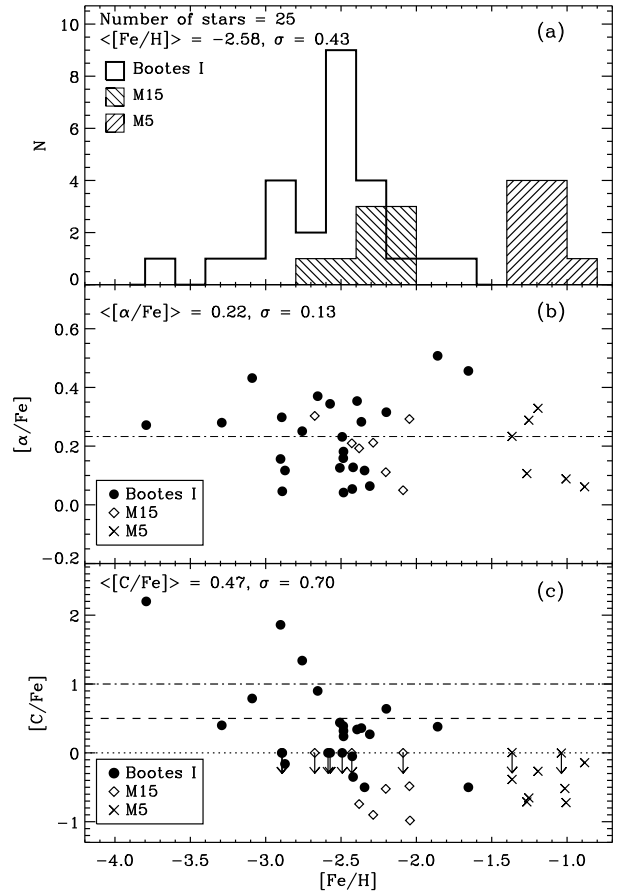}}
\end{center}
\figcaption{(a) The MDF for our
  sample using a bin size of 0.2 dex. The open histogram, along with the
  reported sample values, are for our \bootes{} sample. The filled 
  histograms at [Fe/H]$\sim-2.4$ and [Fe/H]$\sim-1.2$ are the values we measured
  for our M15 and M5 stars, respectively. The metallicity spread and
  metal-poor nature of \bootes{} is contrasted to the M15 and M5
  distributions. (b) The [$\alpha$/Fe] values measured
  for our stars. The filled circles are for the \bootes{} stars, the diamonds
  are for the M15 stars, and the x symbols are for the M5 stars. In
  this case, the distribution of [$\alpha$/Fe] is very similar between
  \bootes{} and the cluster stars. (c) The [C/Fe]
  distribution of our sample. The [C/Fe] for \bootes{} range is both
  clearly  larger and higher than the [C/Fe] of the cluster stars.
\label{summary}}
\end{figure}

Table \ref{infotable} presents the [Fe/H], [$\alpha$/Fe], and [C/Fe]
abundance results for the \bootes{} sample. Overall we find an average
[Fe/H]$=-2.59$, and $\sigma([Fe/H])=0.43$ dex. To take into account
the estimated errors on [Fe/H] and calculate an intrinsic spread of
[Fe/H] we use Equation 8 from \citet{kirby11a}. The intrinsic spread
we find in our sample is $\sigma([Fe/H])=0.38$. Figure
\ref{summary}(a) shows the metallicity distribution function (MDF) of
our \bootes{} sample. It is apparent that the \bootes{} MDF has a wide
range of metallicities (2.1 dex in [Fe/H]), but is sharply peaked at
the average metallicity of [Fe/H]$=-2.59$. Also plotted in Figure
\ref{summary}(a) are our measurements of the MDF in our M5 and M15
sample. The spread at their respective metallicities is an indication
of the error on [Fe/H] from this study.

One star in our sample, Boo21, stands out as exceptionally metal poor,
[Fe/H]$=-3.79$. This is one of the more Fe-poor stars known in any
dSph galaxy, matching the [Fe/H]$=-3.81$ star found in the Sculptor
dSph \citep{frebel10b}.  This is also the second extremely Fe-poor
star known in \bootes{}. A [Fe/H]$=-3.66$ star was discovered by
\citet{norris08} and subsequently verified by \citet{norris10}.  Given
the rather low number of stars with [Fe/H]$<-3.5$ presently known in
the Galactic halo, these results suggest that the ultra-faint dSph
like systems were an important source of stars in this low range of
metallicity.

\subsection{The \bootes{} metallicity distribution}

The detailed properties of the \bootes{} metallicity distribution are
important constraints for its past evolution. For example, where the
$\langle {\rm [Fe/H]} \rangle$ of \bootes{} falls on the
metallacity-luminosity relationship can be indicitave of whether dwarf
galaxies follow evolutionary trends of their more luminous
counterparts, or are actually stripped versions of these luminous
counterparts. Additionaly, measuring its intrinsic metallicity spread
can inform the nature of its chemical enrichment and duration of star
formation. 

Finally the actual shape of the MDF can be also be used to
more carefully constrain its chemical evolution history. For example,
effective yield could decrease with decreasing stellar mass because fo
the increasing importance of outflows and supernova
feedback. However, \citet{prantzos08} assumes that the effective yield in lower
mass systems (i.e. accreted satellites that formed the halo) should
have even lower values than what would be predicted from more massive
systems because of the lack of Type Ia contributions in these
systems. \bootes{} is a low stellar mass system which shows no sign of
Type Ia contribution with its constant [$\alpha$/Fe] (see Section \ref{alpha}),
indicating that its star formation history is truncated enough to test
this idea.   

\subsubsection{Average and Intrinsic Spread in Metallicity}

To get a more detailed understanding of these properties, we expand
the our \bootes{} [Fe/H] sample to include the non-overlapping stars
from \citet{norris10b} (with [Fe/H] taken from \citet{feltzing09} if
a high-resolution derivation was not available), and compare to the
dSph findings of \citet{kirby11a}. In the case of $\langle {\rm
  [Fe/H]} \rangle$ here, we employ a weighted average using the inverse
square of the errors because of the different methods used to
determine metallicity. For simplicity, we assume [Fe/H] errors of 0.1,
0.2, and 0.3 dex, for high-resolution determinations, our present
determinations, and the Ca II K determinations, respectively.

For this expanded sample of 41 stars we find $\langle {\rm [Fe/H]}
\rangle = -2.51$ and intrinsic spread of $\sigma{\rm
  [Fe/H]}=0.40$. The $\langle {\rm [Fe/H]} \rangle$ value is very
close to the \citet{norris10b} result, and as such seems to extends
the metallicity-luminosity relationship in this regime. In the sense
that \bootes{} is representative of a normal galaxy, one natural
explanation for this low metallicity is that the effective yield
decreases with decreasing stellar mass, reflecting the increasing
importance of outflows from supernova feedback in lower mass galaxies
(e.g., \citealt{tremonti04,prantzos08}).

In more detail, we can compare $\langle {\rm [Fe/H]} \rangle$ with the
relationships derived for dSphs by \citet{kirby11a} (their Equation
10) and for Local Group dwarf galaxies by \citet{woo08} (as recast by
\citealt{kirby11a} in their Equation 12). Using the \bootes{}
luminosity derived by \citet{martin08}, the metallicities predicted by
these relationships are [Fe/H]$=-2.27$ and [Fe/H]$=-2.18$ for the
\citet{woo08} and \citet{kirby11a} samples,
respectively. Interestingly, our $\langle {\rm [Fe/H]} \rangle$ is
closer to the relationship derived from the \citet{woo08} sample that
uses no ultra-faint dSphs.

The intrinic spread of metallicity of our expanded sample is also very
similar to the value found by \citet{norris10b} for their \bootes{}
stars. The intrinsic spread is in the lower end of the ultra-faint
dSph distribution \citep{norris10b,kirby11a}. By itself, this may
indicate a less stochastic chemical enrichment history, more in line
with the classic dSphs than the ultra-faint dSphs. An alternative way
to understanding the intrinsic spread in metallicity can be had by
converting [Fe/H] to Z, and comparing intrinsic
log$\sigma(Z/Z_{\odot})$ values \citep{kirby11a}. Using this metric,
we find log$\sigma(Z/Z_{\odot})=-2.45$. This is much lower than the
log$\sigma(Z/Z_{\odot})=-2.04$ that would be predicted from the
relationship with luminosity that \citet{kirby11a} derives from other
dSphs. This could be from having a relatively short star formation
duration caused by an inability to retain gas, even when compared to
other ultra-faint dSphs.

\subsubsection{The \bootes{} MDF}

\begin{figure}
\begin{center}
\scalebox{1.15}[1.15]{
\plotone{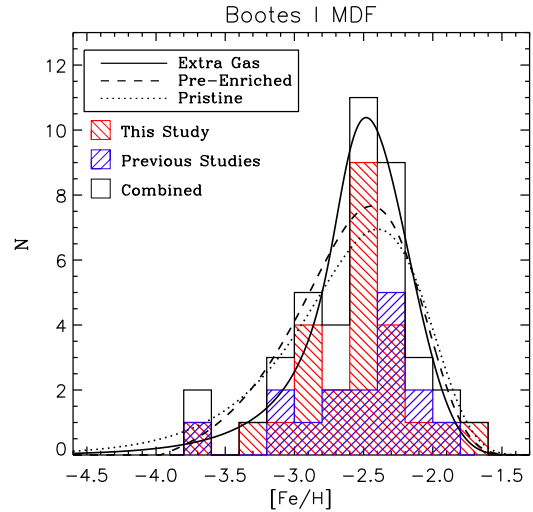}}
\end{center}
\figcaption{ The MDF of \bootes{}, including results from this study 
and previous results reported from \citet{feltzing09} and
  \citet{norris10b}. The various curves represent the best-fit
  analytic chemical evolution models, as described in
  \citet{kirby11a}. 
\label{mdf}}
\end{figure}

To get a more detailed understanding of the \bootes{} star formation
history, we fit the combined MDF from the expanded sample using the simple analytic chemical
evolution models and maximum likelihood fitting technique as described in
\citet{kirby09,kirby11a} (shown in Figure \ref{mdf}). The models
include two leaky-box models, one of which is assumed to start with
metal-free gas (the Pristine Model) and one with pre-enriched initial
gas (the Pre-Enriched Model). We also fit their Extra Gas Model, which
assumes that some new form of gas becomes available for star formation
through either infall or an extra internal source of cooling gas, as
adopted from \citet{lb75}.

In all three models the effective nucleosynthetic yield, $p$, is a
free parameter. This parameter encompasses both metals produced by
supernova that are released back into the ISM as well as gas loss from the
system. The Pre-Enriched Model also has an extra free parameter, the
initial metallicity of the gas, designated by [Fe/H]$_{\rm o}$. For
simplicity the Extra Gas Model assumes that the additional gas has
zero metallicity. The amount of extra gas, $M$ (where $M=1$ represents
no extra gas and therefore reduces to the Pristine Model), is the
extra free parameter.

The results of performing the maximum likelihood fit to each model are
shown as the curves in Figure \ref{mdf}. The best Pre-Enriched model
has a [Fe/H]$_{\rm o}=-3.97$. The best Extra Gas model has $M=6.4$;
i.e., 84\% of the stars formed from an extra reservoir of metal-free
gas. These final models can be compared by the logarithm of the ratio
of the maximum likelihoods $L_{\rm max}$ \citep{kirby11a}. Using this
description, the fits give ${\rm ln} (L_{\rm max}({\rm{Extra
    Gas)})}/L_{\rm max}({\rm Pristine}))=0.17$ and ${\rm ln} (L_{\rm
  max}({\rm{Pre-Enriched}) }/L_{\rm max}({\rm Pristine}))=0.07$. So
while the Extra Gas and Pre-Enriched models both fit the distribution
better than the Pristine model, all three actually fit the MDF
reasonably well. 

The very low [Fe/H]$_{\rm o}=-3.97$ for the
Pre-Enriched model shows that assuming pristine gas for the simple
model is an excellent assumption in the case of \bootes{}, and there
is very little difference in the model fits as a result. This is in
contrast to the dSphs studied by \citet{kirby11a}, with only Sculptor
coming close to this low of a [Fe/H]$_{\rm o}$. Qualitatively, the
Pristine and Pre-Enriched model are very similar, while the Extra Gas
model does the best at fitting the peak of the MDF. The Extra Gas
model improves the fit to the peak of the MDF by mainataining a low
metallicity for period of time in the galaxy thanks to the infall of
pristine gas. However, it does a poorer job of fitting the
low-metallicity tail.

Not suprisingly, given the similarity of the fits, the effective
yields are also similar. The $p$ values for the Pristine Model, the
Pre-Enriched Model, and the Extra Gas Model are 0.0040, 0.0035, and
0.0033, respectively. These low and roughly comparable values in $p$
are in agreement with the idea that the effective yield decreases with
mass of the galaxy, through a mechanism such as supernova feedback
having a greater affect on the gas of the gas due to heating or mass
loss. In particular, the $p$ value in \bootes{} continues the monotonic decrease
with decreasing luminosity as found in classic dSphs studied by
\citet{kirby11a}. However, comparing the predicted $p$ value from the
relationship assumed by \citet{prantzos08} by their Equation 4, our
most likely $p$ values are a factor of two higher (assuming the
stellar mass measured by \citealt{martin08}). At least for the case of
\bootes{}, the effective yield is actually better represented by the
relationship defined by more massive local group galaxies
\citep{dw03}.

\subsection{The $[\alpha/Fe]$ distribution \label{alpha}}

Figure \ref{summary}(b) shows a very similar distribution of
[$\alpha$/Fe] for the \bootes{} stars as for the cluster stars. As
this is a relative comparison, this takes into account systematic and
random errors. The average and standard deviation for all of the
\bootes{} stars are $\langle$[$\alpha$/Fe]$\rangle=0.23$ and
$\sigma=0.14$ dex. For M5 these measurements are
$\langle$[$\alpha$/Fe]$\rangle=0.18$ and $\sigma=0.11$ dex, and for M15,
$\langle$[$\alpha$/Fe]$\rangle=0.20$ and $\sigma=0.09$ dex. Given the
error in measuring [$\alpha$/Fe], the \bootes{} distribution is
consistent with the measured cluster distributions. In particular,
because both M5 and M15 show no signs of [$\alpha$/Fe] variations in
high-resolution studies, our \bootes{} sample is consistent with
showing no spread in [$\alpha$/Fe].

Since it is well known that the globular cluster and halo [Fe/H]
distributions are very similar, we conclude that, on average, the
\bootes{} stars are also a good match to the halo [$\alpha$/Fe]. We
find a flat distribution of [$\alpha$/Fe] in the metallicity range of
our \bootes{} stars, which is typically attributed to having mainly
contributions from Type II SN, and none to very little from Type Ia
events. This flat trend is in contrast to the nearly constant decline
of [$\alpha$/Fe] found in classic dSphs by \citet{kirby11b}. However,
as they note, their data are sparse for [Fe/H]$<-2.5$, where most of
the \bootes{} stars lie.

We interpret this flat trend in a similar manner to halo metal-poor
stars, an indication that Type Ia SN have had little to no contribution to
the \bootes{} chemical composition. If we assume we have a systematic
offset in our [$\alpha$/Fe] measurements (section 3.1), we estimate
that the true value for $\langle$[$\alpha$/Fe]$\rangle$ in \bootes{}
is 0.36.

\subsection{[C/Fe] in \bootes{}}

Figure \ref{summary}(c) shows the distribution of [C/Fe] for our
sample. While the cluster stars show a clear subsolar average in
[C/Fe], the \bootes{} sample has a much larger spread, which includes
many carbon-enhanced metal-poor (CEMP) stars. This is typically
defined by stars with [Fe/H]$\le-2.0$ and [C/Fe]$\ge1.0$. Using this
traditional definition (but relaxing the [Fe/H] constraint to include
the two stars with slightly higher metallicities in this particular
\bootes{} data set), the CEMP fraction of this sample is 12\% $\pm$
7\%. Even with the relatively large error due to small numbers, this
is in line with estimates of the halo fraction of CEMP stars that
range from 9 to 25 \% (see, e.g. \citealt{cohen05, lucatello06,
  lai07}). One of these CEMP stars includes the most Fe-poor star of
our sample, Boo21, which has [C/Fe]$=+2.2$. The metallicity and carbon
enhancment for this star matches the measurements of another dSph star
found by \citet{norris10c} in the Segue 1 dSph system. However, the
resolution of the spectrum is too low to determine if it is a CEMP-no
star (carbon-rich, metal-poor, and not enhanced in neutron-capture
elements) like the Segue 1 star.

Our carbon-enhanced star fractions are 
in contrast with the sample of \citet{norris10b}, where no CEMP stars
were discovered in \bootes{}. Including these stars in the calculation,
the CEMP fraction drops to 8\% $\pm$ 4\%. However, even this lower
figure is still consistent with at least the lower bound of the
estimates CEMP fractions measured in the halo.

Alternative definitions for CEMP stars has been proposed by
\citet{aoki07} and by \citet{norris10b}. The \citet{aoki07} definition
is dependent on luminosity and takes into account an expected reduction of
[C/Fe] due to mixing events as a star evolves up the red giant branch.
\citet{norris10b} suggests a definition of carbon richness
that is dependent on [Fe/H], utilizing the same set of data presented
in \citet{aoki07}.

\begin{figure*}
\begin{center}
\scalebox{1.}[1.]{
\plotone{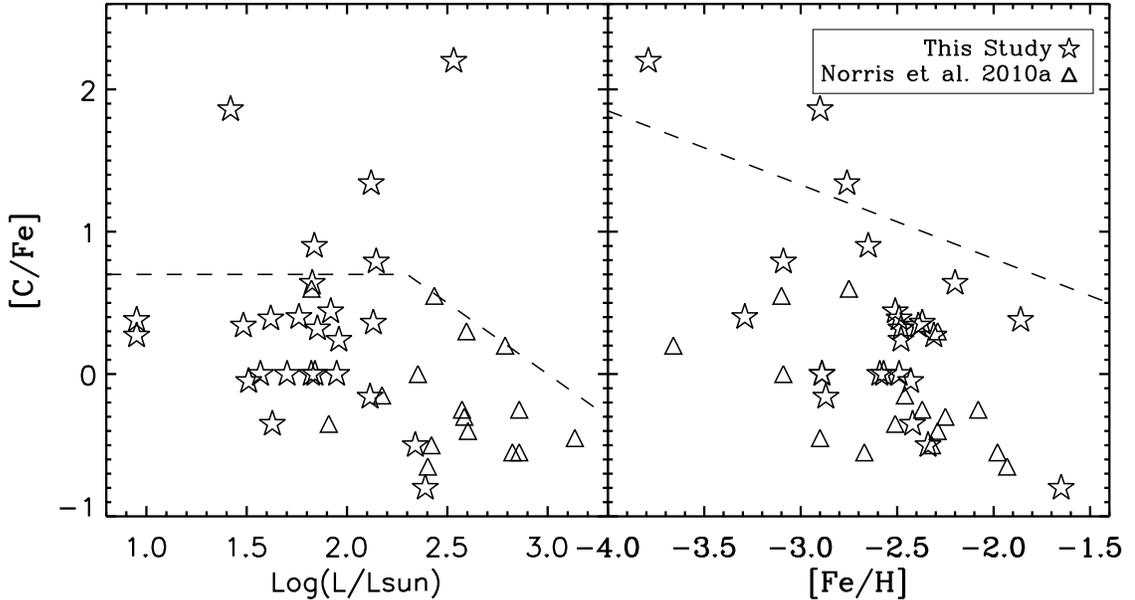}}
\end{center}
\figcaption{ [C/Fe] of \bootes{} stars as a function of Log($L/L_{\odot}$) and
  [Fe/H]. The dashed lines show the lower limit for carbon richness as
  defined by \citet{aoki07} in the Log($L/L_{\odot}$) plot, and by
  \citet{norris10b} in the [Fe/H] plot. 
\label{crich}}
\end{figure*}

Both of these situations are plotted in Figure \ref{crich}. In the
luminosity dependent case, the CEMP fraction increases to 20\% $\pm$
9\% and 12\% $\pm$ 5\%, for this study's data and for this study's
combined with the \citet{norris10b} study's results, respectively. For
the [Fe/H] dependent definition, the percentages remain the same with
respect to the traditional CEMP definition. So regardless of the
definition adopted, the CEMP fraction of \bootes{} appears consistent
with the halo fraction.

\section{Discussion}

\bootes{} is a very metal poor system; the average $\langle {\rm
  [Fe/H]} \rangle = -2.59$ found here agrees with previous studies of
this system. We have discovered a star with [Fe/H]$=-3.79$, which is
essentially identical to lowest discovered in any dSph system, either
classic or ultra-faint, and is the second such star discovered in
\bootes{}. There is a large total range in [Fe/H] of 2.1 dex. This is
larger than the range found by \citet{norris10b}, 1.7 dex, but the two
values are consistent within errors. In particular, the intrinsic
spread in metallicity $\sigma({\rm [Fe/H]})=0.38$, is almost identical
to the value measured by \citet{norris10b}.

Combining the MDF from this study with previously measured values and
comparing them to simple analytic chemical evolution models indicates
that the stellar content of \bootes{} may be dominated by stars formed
from a fresh supply of gas distinct from the initial star forming gas
(either externally through infall/accretion or an internal gas that
cools between episodes of star formation). However, leaky-box models
assuming initially zero-metallicity and pre-enriched gas also fit the
MDF comparably well. Consistently from all three models, we
find \bootes{} has an effective yield that follows the monotonically
decreasing trend of effective yield with lumonisity as found in the
classic dSphs measured by \citet{kirby11a}. The actual value,
$p \simeq 0.0035$, is about a factor of two higher than the trend
assumed by \citet{prantzos08} when working with assumption of no Type
Ia SN contributions. 

Overall, the bulk metallicity properties ($\langle {\rm
  [Fe/H]} \rangle$, $\sigma({\rm [Fe/H]})$, and $p$) of \bootes{} point to a
system that falls in more in line with trends from more luminous
counterparts \citep{woo08,kirby11a}. However, given the general spread
found in these quantities found for other ultra-fain dSph galaxies,
\bootes{} is not so much of an outlier, but is likely more
of an indication that even at these low luminosities there can be
appreciable variations in chemical evolution histories.

We also find that, on average, our [$\alpha$/Fe] matches the halo
pattern at the same metallicites, showing no signs of Type Ia SN
contributions. The CEMP fraction we calculate, 12 \% for just this
sample and 8 \% including the \citet{norris10b} results, also matches
the halo CEMP fraction. These combined results place \bootes{} in a
very interesting position. On {\it average} the abundance patterns we
find are good matches to those seen in metal-poor halo stars in the
same low metallicity regime of \bootes{} ([Fe/H] $\approx -2.0$ and
lower). However, \citet{feltzing09}
finds variations in [Mg/Ca] in \bootes{}, which match similar findings
in other dSphs Hercules \citep{koch08} and Draco \citep{fulbright04}
for stars that are in the same low metallicity range as defined by
\bootes{} stars. Given the rarity of finding this type of abundance
signature in the halo, this might indicate a disimilarity of
populations between low metallicity stars in the halo and dSphs.

Furthermore, careful comparison of the MDF of the Sculptor dSph by
\citet{kirby09} with the metal-poor tail of the halo field from
\citet{schorck09} also demonstrates another possible disconnect
between metal-poor dSph populations and the halo field; too many
extremely metal-poor (EMP; [Fe/H] $< -3.0$ stars), relative to the
halo, being found in Sculptor. On the other hand, this study finds a
CEMP fraction similar to that found in the halo field. Also, others
have found that abundance patterns for some stars in both the
ultra-faint and classic dSph systems match the detailed abundance
patterns of halo EMP stars \citep{frebel10a,frebel10b, norris10c}.

As has been previously pointed out (e.g., \citealt{kirby09}), some of
these disimilarities should actually be expected from formation
models (e.g, \citealt{robertson05,font06}), where only a
few dwarf systems more massive than current day dSph systems formed
the majority of the halo population. This is further supported by the
differing $\alpha$-element abundance patterns found in dSph stars with
metallicitities higher than those found in \bootes{} (see 
\citealt{geisler07} and references therein).

However, this is true for the inner-halo population. Farther out into
the halo, there is evidence for a dichotomy in the populations
\citep{carollo07,carollo10}. This may be because the history of the
outer halo region is more dominated by recent accretion and smaller
mass systems \citep{bullock05}. In this same regard, \citet{carollo11}
have recently shown that the fraction of CEMP stars at a given (low)
metallicity that can be kinematically associated with the outer-halo
population is roughly twice that of the frequency associated with the
inner-halo population, and took this as strong evidence for multiple
additional sources of carbon production beyond the canonical AGB-star
mechanism that may dominate in the inner halo.

Considering that most halo stars that have been studied are members of
the inner-halo population, the diversity of abundance signatures found
in dSph systems compared to what has been found so far in halo stars
is not surprising. The similarities in both average properties as
found here for \bootes{}, and in detailed chemical abundance patterns
found in individual dSph stars with halo stars, do seem to indicate
some commonality in chemical evolution in dSphs with Galactic halo
progenitors. However, at least in the inner-halo population, the more
unique chemical signatures found in some individual dSph stars may be
washed out by stars contributed from the larger progenitor systems
predicted by \citet{robertson05} and \citet{font06}.

Given all of the above, the growing body of abundance results for both
classic and ultra-faint dSph systems such as \bootes{} points to the
outer-halo population of the Galaxy as being a promising hunting
ground for validation of the hierarchical formation scenario as
described in \citet{bullock05}. There is already evidence of a more
diverse outer halo, in both chemically unique individual stars
\citep{lai09} and in chemical diversity as a whole
\citep{roederer09}. Further tests of this picture will require
measuring abundances for both a significantly larger number of dSph
stars and for additional in situ studies of the outer halo.

\acknowledgements

We thank the anonymous referee for their useful comments and suggestions.

D.K.L acknowledges the support of the National Science Foundation through the NSF Astronomy
and Astrophysics Postdoctoral Fellowship under award AST-0802292.

Y.S.L. and T.C.B acknowledge partial support for this work from the NSF under
grants PHY 02-16783 and PHY 08-22648; Physics Frontier Center/Joint Institute
for Nuclear Astrophysics (JINA).

S.L. is grateful to the DFG cluster of excellence ''Origin and Structure of the Universe'' for partial support.

M. B. and J. A. J. acknowledges support from NSF grant AST-0607770 and
AST-0607482.

{\it Facilities:} \facility{Keck:I (LRIS)}

\bibliography{/Users/david/Dropbox/my_papers/bibfile/all.bib}

\clearpage

\end{document}